\documentclass{iopart}
\usepackage[utf8]{inputenc}
\usepackage[T1]{fontenc}
\usepackage[english]{babel}
\usepackage[colorlinks,citecolor=blue]{hyperref}
\usepackage{times}
\usepackage[sort&compress,numbers]{natbib}
\usepackage{graphicx}

\begin{document}
\title{Adaptive-network models of swarm dynamics}
\author{Cristián Huepe$^1$, Gerd Zschaler$^2$, Anne-Ly Do$^2$ and Thilo Gross$^2$}
  \address{$^1$614 N. Paulina Street, Chicago IL 60622-6062, USA}
  \address{$^2$Max-Planck-Institut für Physik komplexer Systeme, Nöthnitzer Str. 38, 01187 Dresden, Germany}
\ead{cristian@northwestern.edu}

\begin{abstract}
We propose a simple adaptive-network model describing recent swarming experiments.
Exploiting an analogy with human decision making, we capture the dynamics of the model
by a low-dimensional system of equations permitting analytical investigation.
We find that the model reproduces several characteristic features of swarms, including spontaneous
symmetry breaking, noise- and density-driven order-disorder transitions that can be of
first or second order, and intermittency.
Reproducing these experimental observations using a non-spatial model suggests that spatial geometry may have a lesser impact on collective motion than previously thought.
\end{abstract}

\pacs{87.10.-e, 64.60.Cn, 64.60.aq}
\submitto{\NJP}

\maketitle

\section{Introduction}
``More is different,'' a central observation in complex systems research is in few areas as evident
as in collective decision making. Recent studies on groups of self-propelled agents, such as bird
flocks, fish schools, insect swarms, or herds of quadrupeds show that these can often make better choices in groups than individually \cite{IainFish,IainCollectiveMinds,CuckerHuepe}.
Effective collective action, which has given rise to the popular notion of {\it swarm intelligence},
appears to follow from universal organizing principles \cite{SwarmIntelligenceBook}.
The ease and elegance with which, e.g., a school of fish finds its way and avoids predators, can be even more appreciated if one compares it for instance to political decision making in humans.

It is interesting to note that, although regarded as collective decision-making processes, swarming behaviour and collective motion are modelled differently from decision making in human populations. Theoretical studies of collective motion have mostly focused on agent-based simulations of self-propelled particles \cite{VicsekCzirokBenJacob, ChateGinelliGregoire, RomanczukCouzin, MachSchweitzer, CristianNewTools, Sumpter2008} and fluid-like models that treat them as continuous media \cite{TonerTu, TopazBertozzi}. By contrast, studies of decision making and opinion formation in social systems typically represent the system as a network, emphasizing the discrete nature of interactions \cite{Holley1975, Liggett1999, Sood2005}. One of the reasons for this difference in modelling approaches is that for swarm systems, spatial embedding is assumed to be of central importance, whereas social interactions are felt to be less constrained by physical space.
Furthermore, in collective motion, it cannot be neglected that an agent's decision to move in a certain direction determines the agents with whom it will be interacting next. However, a similar feedback of individual decisions on future interaction partners was also studied in recent works on opinion formation \cite{Holme2006, GilZanette, VazquezEguiluzSanMiguel, NardiniKozmaBarrat, BenczikEtAl, Kimura, Boehme2011, Demirel2011}. The resulting models incorporate both an opinion formation process on the network and a dynamic update of the network topology and thus fall into the class of adaptive networks \cite{ThiloANReview, ThiloANBook}.

In the present article, we propose a non-spatial adaptive-network (AN) model of swarming behaviour.
Reproducing characteristic observations for swarm systems, we find that spatial geometry might play a less central role than has been assumed. Our approach highlights the analogy between swarming and social consensus, thus building a bridge between two areas of research that have been so far considered separately.

\section{Adaptive-network model for a swarming experiment}
We focus on the swarming experiments by Buhl et al. \cite{IainLocusts}.
In their set-up, groups of 5 to 120 locusts were
placed in a ring-shaped arena and left to march freely for 8 hours
while a digital camera captured their positions and orientations. At
low insect number, no ordered collective motion arises; the system
displays no clear clockwise or counter-clockwise flow of locusts around
the arena. At intermediate insect numbers, locusts start aligning,
generating long periods of collective rotational motion during
which most agents are marching in the same direction. These periods of
coherent motion are interrupted by rapid spontaneous changes in their
collective heading direction. Finally, at high insect number this spontaneous direction
switching is no longer observed and agents rapidly adopt a common
and persistent marching direction, either clockwise or counter-clockwise.
Buhl et al.\ reproduced these experimental results qualitatively in
simulations using a one-dimensional agent-based model of self-propelled
particles, also investigating the effect of inherent noise in a subsequent study \cite{IainLocusts, Yates2009}.

Here, we model this experiment using a different approach, trying to address the mechanism leading to the observed collective dynamics with the help of a simple low-dimensional description that lends itself to analytical treatment. To that end, we consider the system of interacting agents as a complex network.
Each node represents an insect, and nodes are linked if the corresponding locusts are
mutually aware of each other through any interaction mechanism.
As in \cite{IainLocusts}, we distinguish only two directions of motion:
every node can be in an R or L state, representing an agent that
marches clockwise (a {\it right-goer}) or counter-clockwise (a {\it left-goer}), respectively.
We refer below to pairs of nodes (agents) in the same state as {\it equal-goers} and to
those in different states as {\it opposite-goers}.

The proposed model only takes into account the agents' headings and contact
network while neglecting all other information, including insect positions.
The evolution of the network is modelled by a set of stochastic processes.
As locusts advance in the experimental system, non-interacting opposite-goers eventually meet
and start sensing each other.
We model this by randomly introducing R-L links at a rate of $a_o$ per node.
In addition, interacting opposite-goers will eventually lose contact,
which is modelled by the random deletion of R-L links at a rate of $d_o$ per link.
Likewise, equal-goers can start or stop interacting as they approach or separate
from each other due to marching speed differences or lateral displacements.
This is represented by also introducing for equal-goers the attachment rate per node $a_e$
and deletion rate per link $d_e$.
Using these conventions, all rates are defined as intensive quantities.
The state dynamics of each node is given by a stochastic process
that depends on its topological neighbours.
We assume that each node switches direction with probability $w_2$ for every R-L link it has
to an opposite-goer.
To account for non-linear three-agent interactions, we introduce an additional
probability $w_3$ of the central node switching direction for every L-R-L and R-L-R chain.
Finally, noise is represented by a constant probability $q$ of an agent
spontaneously switching direction.

\section{Analytical description}
In order to study the collective dynamics of the AN model, we define convenient observables,
the so-called {\it moments}, given by the densities of various subgraphs in the network
\cite{KeelingRandMorris, ThiloEpidemics}.
Each subgraph can be classified by its order, i.e.\ the number of links it contains.
Zeroth order moments are given by the right- and left-goer densities ($[R]$ and $[L]$, respectively).
First order moments are the per-capita densities of R-R, R-L,
and L-L links ($[RR]$, $[RL]$, and $[LL]$).
Second order moments correspond to the densities of A-B-C triplets $[ABC]$,
with $A, B, C \in \{ R, L \}$.
The moment dynamics is captured by balance equations containing the variables of
interest together with densities of larger subgraphs.

The zeroth-moment equations are
\begin{equation}
\label{eqn:zerothorder}
\frac{{\rm d}}{{\rm dt}}[R]  = q \left( [L] - [R] \right) + w_3 \left( [RLR] -[LRL] \right),
\end{equation}
and the symmetric expression for $[L]$, obtained by interchanging $R$ and $L$.
The first-moment equations are
\begin{eqnarray}
\label{eqn:[RR]}
&\frac{{\rm d}}{{\rm dt}}[RR] =  q \left( [LR] - 2 [RR] \right) + w_2 \left( [LR] + 2 [RLR] - [RRL] \right) \nonumber \\
&    + w_3 \left( 2 [RLR] + 3 [^RL^R_R] - [^RR^L_L] \right) + a_e [R]^2 - d_e [RR],
\end{eqnarray}
and the symmetric expression for $[LL]$.
Here we use $[^AB^C_D]$ to denote
the density of third-order motifs with a central node in state $B\in \{ L,R \}$ linked to three nodes in
states $A,C,D \in \{ L,R \}$.
Finally, rather than writing an equation for $[LR]$, we note that the {\it total} first-moment link dynamics
depends only on the link creation and deletion processes through
\begin{eqnarray} \label{eqn:LR}
{\rm \frac{d}{dt}} \left( [LR] + [RR] + [LL] \right) =
    a_o [L][R] - d_o [LR] + \nonumber \\
    a_e \left( [R]^2 + [L]^2 \right) - d_e \left( [LL] + [RR] \right).
\end{eqnarray}
The ODE system \eref{eqn:zerothorder}--\eref{eqn:LR} can now be closed using a
{\it pair-approximation} \cite{KeelingRandMorris,ThiloEpidemics,Zschaler2010}, where
triplets and quadruplets are given by
\begin{equation}
[RLR] = \kappa \frac{[LR]^2}{2 [L]} \ , \quad
[RRL] = 2\kappa \frac{[LR] [RR]}{[R]}, \nonumber
\end{equation}
\begin{equation}
[^RL^R_R] = \kappa^2 \frac{[LR]^3}{6 [L]^2} \ , \quad
[^RR^L_L] = \kappa^2 \frac{[LR]^2 [RR]}{[R]^2}, \nonumber
\end{equation}
and symmetric expressions.
The factor $\kappa = \left(\langle k^2 \rangle - \langle k \rangle\right)/\langle k \rangle^2$ relates the second and first moments of the degree distribution.
Because our network dynamics will yield an unknown, randomly evolving topology, we
use a random graph approximation, setting $\kappa = 1$ as in
\cite{VazquezEguiluzSanMiguel,KeelingRandMorris,ThiloEpidemics,Kimura,Zschaler2010}.
\begin{figure}\centering
\includegraphics[width=9cm]{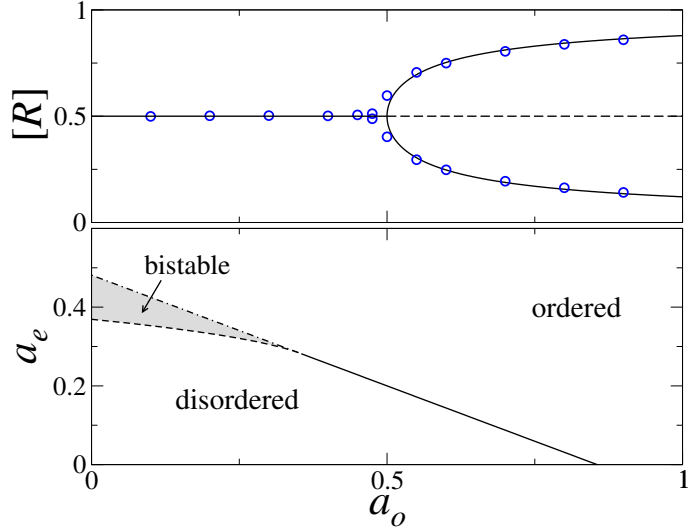}%
\caption{\label{fig1}
Top: Bifurcation diagram of the density of right-goers $[R]$
vs.\ link creation rate $a_o$.
Solutions of the ODE system of Eqs.\ (1)-(3) (solid line) yield
a supercritical pitchfork bifurcation in excellent agreement
with results from numerical network simulations (circles).
Bottom: Phase diagram showing the bifurcation point as a function
of the link creation rates $a_o$ and $a_e$.
In the bistable region (grey), the pitchfork bifurcation becomes subcritical.
Parameters: $N=10^4$ nodes, $d_o = 0.25$, $d_e = 0.1$,
$ w_2 = w_3 = 0.2 $, $q = 0.1$, and (top only) $a_e = 0.2$.
}
\end{figure}

For $a_e = d_e = 0$, the stationary solutions of this ODE system decouple,
with Eqs.\ \eref{eqn:zerothorder} and \eref{eqn:LR} solved independently.
We obtain analytically a mixed-phase solution branch ($[R]=[L]=1/2$) that
becomes unstable in a supercritical pitchfork bifurcation at
$a_o^* = 2 d_o \sqrt{2 q / (\kappa w_3)}$,
giving rise to the two collective-motion solution branches
$ [R]_{\pm} = (1/2) \pm \sqrt{1 - 8 q d_o^2 / (\kappa w_3 a_o^2) } / 2$.
For $a_e \neq 0$ and $d_e \neq 0$, the stationary solutions can be computed
numerically by solving the corresponding system of algebraic equations.
Here, we also find a supercritical pitchfork bifurcation for small $a_e$,
as shown on \fref{fig1} (top).
But for higher values of $a_e$, the transition occurs through a subcritical pitchfork bifurcation
(\fref{fig2}, right-column insets).
This yields a bistable phase where ordered and disordered states coexist,
highlighted in \fref{fig1} (bottom).

We note that a supercritical pitchfork bifurcation, given by $[R]_{\pm} = (1/2) \pm \sqrt{ 1 - 4 q / w_3 } / 2$, could be already observed in a simpler approximation, in which the system is closed at the zeroth order. However, the pair approximation is more accurate when compared with stochastic simulations of the network. Moreover, it allows for the new class of subcritical solutions, which we discuss below.

\section{Results}
\begin{figure}\centering
\includegraphics[width=9cm]{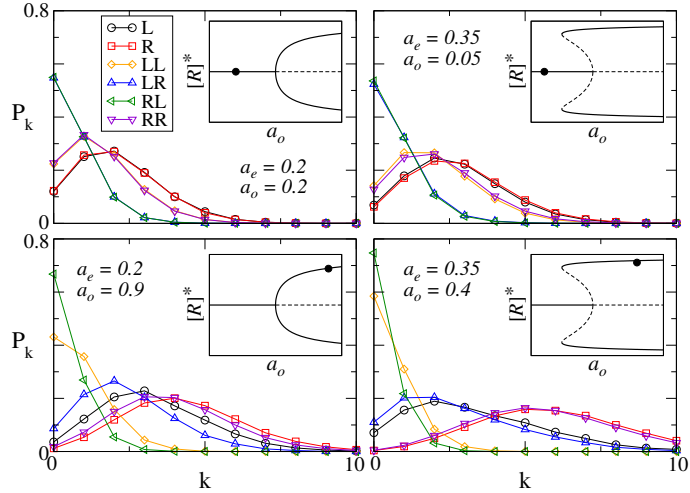}
\caption{ \label{fig2}
Degree distributions of four different stationary solutions
obtained through adaptive network simulations.
The top (bottom) row shows cases in the disordered
(ordered) phase, with insets displaying their location in the
bifurcation diagram.
The left (right) column shows cases with a supercritical
(subcritical) pitchfork bifurcation.
Different curves display the connectivity of left-goers (L),
of right-goers (R), of left-goers only to left-goers (LL) or only to
right-goers (LR), etc.
Parameters: Same as in \fref{fig1}, except when noted on plots.
}
\end{figure}

In \fref{fig1} (top), we show that the ODE system solutions are in excellent agreement with individual-based stochastic simulations of the network dynamics. We verified that the small remaining discrepancy is due to the $\kappa=1$ assumption, and not other factors such as finite-size effects.

We now compare the results of our AN model to the locust
experiments in \cite{IainLocusts}.
The bifurcation diagram in \fref{fig1} uses the encounter rate
between opposite-goers $a_o$ as control parameter, which
is proportional to the experimental agent density in the current framework.
We can thus compare our results directly to the
behaviour observed experimentally at different agent numbers.
For a low association rate $a_o$, the network exhibits no order or symmetry breaking.
This corresponds to the disordered mix of left- and right-goers showing no
collective motion obtained in experiments at low insect number.

For high $a_o$, the system must be in one of the two solution branches, with a majority
of nodes in either state (R or L). This is the ordered collective marching state found at high
insect number \cite{IainLocusts}.
A similar pitchfork bifurcation is also observed when using the noise intensity, $q$,
instead of $a_o$ as control parameter, as done in most previous numerical works
\cite{VicsekCzirokBenJacob,ChateGinelliGregoire,CristianNewTools}.

Let us emphasize that the presence of this transition in the AN model implies that very
few elements of the agent dynamics are required to obtain such swarming behaviour.
In particular, we did not choose any specific interaction rule but only required that
it drives agents to head in the same direction.
By contrast, we find that three-body interaction processes are required to break the symmetry
and obtain swarming solutions.
Furthermore, a subcritical bifurcation, giving rise to hysteresis or sudden polarisation, is only possible in the AN model if $a_e$ and $d_e$ are non-zero.
This qualitative result could shed light on the current controversy over the order of the swarming transition
\cite{VicsekCzirokBenJacob,ChateGinelliGregoire,NagyDarukaVicsek,CristianBirds,CristianNet}.
Indeed, a first-order transition (stemming from the subcritical pitchfork bifurcation)
is only possible here if groups of equal-goers can associate or dissociate while
heading in the same direction. It would be very interesting to explore if a similar effect is present
in agent-based simulations and experiments.

\Fref{fig2} shows the degree distributions obtained in the ordered and disordered
phases for supercritical and subcritical cases.
Both display similar connectivities.
In the disordered states (top), most agents have very few links to equal-goers and
no links to opposite-goers.
This is also observed in agent-based simulations and experiments
\cite{IainLocusts,CristianNewTools}, where the disordered regime develops no large clusters
and, therefore, small connectivity.

In the ordered (right-going) state (bottom), R-R links are strongly favoured.
This corresponds to the formation of large right-going groups in the agent-based dynamics.
The number of L-R links also increases, which corresponds to encounters between a
few left-goers and these large right-going clusters.
The typical number of all other links decreases.

\begin{figure}\centering
\includegraphics[width=9cm]{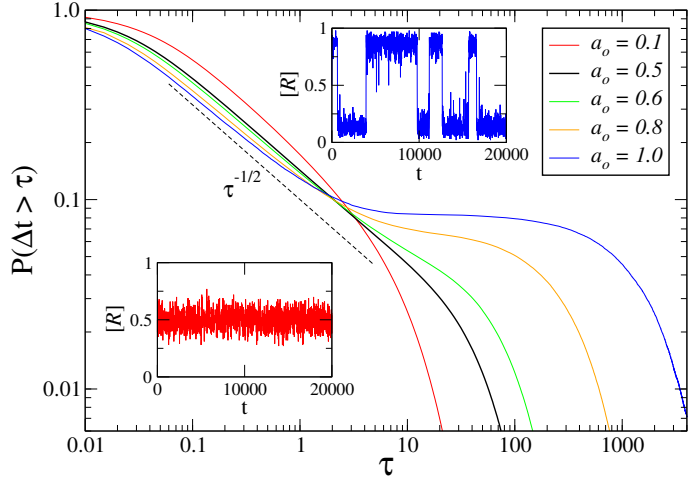}
\caption{ \label{fig3}
Cumulative distribution of residence-times in a majority L or majority R state
for the adaptive network dynamics at various values of $a_o$.
As $a_o$ is increased, the system becomes more ordered, switching
direction less often, and the distribution becomes broader.
For $a_o > 0.6$, a preferred residence time appears at large
$\tau \sim 10^3$ due to finite-size effects.
Insets show the density of right-goers $[R]$ vs.\ time for
$a_o = 0.1$ (bottom-left) and $a_o = 1.0$ (top-centre).
Parameters: Same as in \fref{fig1} (top) but for $100$ nodes.
}
\end{figure}

In previous experiments and simulations it was observed that there is an intermittent regime where the swarm is polarised but can switch the marching direction spontaneously \cite{IainLocusts,CristianIntermit}. This behaviour is also captured by the AN model.
The intermittent regime occurs at intermediate densities,
i.e.\ at $a_o$ values close to the bifurcation, in the ordered phase, where a low
nucleation barrier allows for stochastic switching between the two branches.

\Fref{fig3} shows the cumulative distributions of residence times (lasting
$\tau$ or longer) in which the network resides in a majority R state before
switching to a majority L state or vice-versa.
In the disordered phase, this distribution decays exponentially as expected, since
it results from memory-less stochastic fluctuations about the stationary state.
As $a_o$ approaches its critical value, the distribution
develops a long tail approximating a power law with exponent $-1/2$, providing evidence for a switching process with memory.
We note that a power-law distribution of switching times with the \emph{same} exponent was observed in previous spatial models \cite{CristianIntermit,CristianNewTools}.
The experimental locust dynamics also display intermittency
at intermediate densities, but the available time series are not long enough to
characterize its statistics \cite{IainLocusts}.

For $a_o \geq 0.8$ finite-size effects produce a preferred residence time
at large $\tau \sim 10^3$ that grows with the system size, appearing as a plateau
in the cumulative distributions.
This corresponds to the typical escape time from highly polarised states where the
system gets trapped when system-wide connectivity is reached within the finite network.

\section{Conclusions}
In summary, we have proposed an adaptive-network model of a swarm experiment
that captures much of its characteristic collective behaviour and likens it to an
opinion-formation process.
In particular, our model displays a transition from a disordered to an ordered phase with increasing insect density.
Furthermore, an intermittent regime is observed close to the transition point, where a fat-tailed distribution of residence-times emerges.

We emphasize that in the proposed model, these characteristics of swarming systems are recovered without an explicit spatial representation of the system. This suggests that the spatial context of swarming may not be of central importance for many phenomena.
By contrast, three-body processes and an increased probability of interaction between
agents with intersecting trajectories are found to be essential in our model.

Further work is certainly necessary to test whether the experimentally observed transition is caused by the same mechanism that is at work in our non-spatial model.
In order to improve our understanding of the role of spatial dynamics in swarms,
detailed network- and agent-based simulations should therefore be compared.
We postulate that the same dynamics will be followed by point-like agents, which
are well described by the nodes in our model, but not by spatially extended agents,
where jamming must play an important role.

In the present work, we used a modelling approach originally proposed for social networks. We believe that the analogies we have drawn between swarming phenomena and opinion formation processes could be fruitfully exploited in further studies, thus building a bridge between the two fields.

\ack
The work of CH was supported by National Science Foundation
Grants DMS-0507745 and PHY-0848755.

\providecommand{\newblock}{}

\end{document}